\documentclass[pra,twoside,showpacs,superscriptaddress,twocolumn]{revtex4}

\usepackage{latexsym}
\usepackage[dvips]{graphics}

\begin{document}

\title{How quantum correlations enhance prediction of complementary measurements}

\author{Radim Filip}

\affiliation{Department of Optics, Palack\'y University,
     17.~listopadu 50, 772\,00 Olomouc, Czech~Republic}

\author{Miroslav Gavenda}

\affiliation{Department of Optics, Palack\'y University,
     17.~listopadu 50, 772\,00 Olomouc, Czech~Republic}

\author{Jan Soubusta}

\affiliation{Joint Laboratory of Optics of Palack\'{y} University and
     Institute of Physics of Academy of Sciences of the Czech Republic,
     17. listopadu 50A, 772\,00 Olomouc, Czech Republic}
\affiliation{Department of Optics, Palack\'y University,
     17.~listopadu 50, 772\,00 Olomouc, Czech~Republic}

\author{Anton\'{\i}n \v{C}ernoch}

\affiliation{Department of Optics, Palack\'y University,
     17.~listopadu 50, 772\,00 Olomouc, Czech~Republic}

\author{Miloslav Du\v{s}ek}

\affiliation{Department of Optics, Palack\'y University,
     17.~listopadu 50, 772\,00 Olomouc, Czech~Republic}

\date{\today}

\begin{abstract}
If there are correlations between two qubits then the
results of the measurement on one of them can help to
predict measurement results on the other one. It is an
interesting question what can be predicted about the
results of two complementary projective measurements on the
first qubit. To quantify these predictions the
complementary \emph{knowledge excesses} are used. A
non-trivial constraint restricting them is derived. For any
mixed state and for arbitrary measurements the knowledge
excesses are bounded by a factor depending only on the
maximal violation of Bell's inequalities. This result is
experimentally verified on two-photon Werner states
prepared by means of spontaneous parametric
down-conversion.
\end{abstract}

\maketitle



Immediately after the discovery of quantum mechanics, it
was realized that quantum correlations between two
particles exhibit interesting counterintuitive features
\cite{EPR}. Assuming a pair of maximally entangled qubits
$S$ and $M$, the results of complementary measurements on
qubit $S$ can be, in principle, perfectly predicted from
two appropriate measurements on qubit $M$. Later, it was
shown that quantum mechanics predicts different values of
certain correlations of measurement results than local
realistic theories. Inequalities, which have to be
satisfied within the local realism, were derived by Bell
\cite{Bell64}. The predictions of quantum mechanics were
already satisfactorily experimentally confirmed using pairs
of photons entangled in polarizations
\cite{Kwiat95,Weihs98}. In this Letter, we analyze in
detail how the correlations between the qubits prepared in
a general mixed state enhance our ability to predict the
results of complementary projective measurements on one
qubit when we know the measurement results on the other
one. This enhancement can be described by the quantity that
we will call complementary knowledge excess. We derive a
non-trivial bound on the knowledge excesses which is
determined only by the maximal violation of Bell
inequalities \cite{Horodecki95}. An experimental test of
this restriction on complementary knowledge excesses was
performed using mixed two-photon Werner state prepared by
means of spontaneous parametric down-conversion.

We assume a general mixed state $\rho_{SM}$ of a ``signal''
qubit $S$ and a ``meter'' qubit $M$. Performing two (ideal)
projective measurements $\Pi_{M},\Pi'_{M}$ on qubit $M$,
the prediction of the results of mutually complementary
measurements $\Pi_{S},\Pi'_{S}$ on qubit $S$ can be
improved. Complementarity of measurements on a qubit means
that $\mbox{Tr}\,\Pi_{Si}\Pi'_{Sj}=1/2$ for any $i,j=0,1$
($\Pi_{Si}, \Pi'_{Si}$ are corresponding projectors).
Assuming $\Pi_{S0}=|\Psi\rangle_{S}\langle\Psi|,
\Pi_{S1}=|\Psi^{\bot}\rangle_{S}\langle\Psi^{\bot}|$, we
can expand the state $\rho_{SM}$ in the form
$\rho_{SM}=w|\Psi\rangle_{S}\langle \Psi| \otimes \rho_{M}+
w^{\bot}|\Psi^{\bot}\rangle_{S}\langle\Psi^{\bot}| \otimes
\rho_{M}^{\bot}+
\sqrt{ww^{\bot}}\left(|\Psi\rangle_{S}\langle \Psi^{\bot} |
\otimes \chi_{M}+\mbox{h.c.}\right)$, where $0\leq
w,w^{\bot}\leq 1$, $ w+w^{\bot}=1$ and the meter operators
$\rho_{M},\rho_{M}^{\bot},\chi_{M}$ depend on the choice of
the measurement $\Pi_{S}$. In order to predict the result
of the measurement $\Pi_{S}$ one needs to discriminate
between the mixed states $\rho_{M}$ and $\rho_{M}^{\bot}$
by a projective two-component measurement
$\Pi_{M}=\{\Pi_{M0},\Pi_{M1}\}$ ($\Pi_{M0}+\Pi_{M1}=1$,
$\Pi_{M0}\Pi_{M1}=0$) on the qubit $M$. Using maximum
likelihood estimation strategy, we can guess for each
detection event the most likely result of the measurement
$\Pi_{S}$. Our knowledge can be quantified as the
fractional excess of the right guesses over wrong guesses
in many such experiments repeated under identical
conditions \cite{Englert}. Using our expansion of
$\rho_{SM}$, the total knowledge is ${\bf
K}(\Pi_{M}\rightarrow
\Pi_{S})=\sum_{i}|\mbox{Tr}_{M}\Pi_{Mi}(w\rho_{M}-w^{\bot}\rho_{M}^{\bot})|$,
whereas without the measurement $\Pi_{M}$, the knowledge is
${\bf P}(\Pi_{S})=|w-w^{\bot}|$. The largest value of
knowledge over all $\Pi_{M}$ was introduced as
distinguishability ${\bf
D}(\Pi_{S})=\mbox{Tr}_{M}|(w\rho_{M}-w^{\bot}\rho_{M}^{\bot})|$.
Further, we define a knowledge excess
\begin{equation}
{\bf \Delta K}(\Pi_{M}\rightarrow \Pi_{S}){\bf K}(\Pi_{M}\rightarrow \Pi_{S})-{\bf P}(\Pi_{S}),
\label{excess}
\end{equation}
where $0\leq {\bf \Delta K}(\Pi_{M}\rightarrow \Pi_{S})\leq
1$. It quantifies only that amount of the knowledge which
exceeds the a-priori knowledge ${\bf P}(\Pi_{S})$. The
largest ${\bf \Delta K}(\Pi_{M}\rightarrow \Pi_{S})$ over
all $\Pi_{M}$ can be considered as a distinguishability
excess ${\bf \Delta D}(\Pi_{S})$. Thus $0\leq {\bf \Delta
K}(\Pi_{M}\rightarrow \Pi_{S})\leq {\bf \Delta
D}(\Pi_{S})$. Analogical quantities ${\bf \Delta
K}(\Pi'_{M}\rightarrow \Pi'_{S})$ and ${\bf \Delta
D}(\Pi'_{S})$ can be defined for the complementary
measurement $\Pi'_{S}$.

Intuitively, for a given mixed state $\rho_{SM}$ of a
two-qubit system the knowledge excesses are somehow
restricted by the properties of the state. To derive a
quantitative constraint, we will use the following
expansion of the state: $\rho_{SM}=\frac{1}{4}(\openone
\otimes \openone + \openone \otimes\sum_{l=1}^{3}
m_{l}\sigma_{l}+ \sum_{l=1}^{3} n_{l}\sigma_{l}\otimes
\openone
+\sum_{k,l=1}^{3}t_{kl}\sigma_{k}\otimes\sigma_{l})$, where
$\openone$ stands for the identity operator; $m_{l},n_{l}$
are vectors in $R^{3}$; $\sigma_{l}$, $l=1,2,3$ are the
standard Pauli operators. The coefficients $t_{kl}$ form a
real correlation matrix $T$ and the vectors $m_{l}$ and
$n_{l}$ determine the local states
$\rho_{S}=\frac{1}{2}(\openone + \sum_l n_{l}\sigma_{l})$,
$\rho_{M}=\frac{1}{2}(\openone + \sum_l m_{l}\sigma_{l})$.
There is a subset of the states $\bar{\rho}_{SM}$ having a
diagonal correlation tensor
$\bar{T}=\mbox{diag}(\bar{t}_{33},\bar{t}_{11},\bar{t}_{22})$,
with the following property
$\bar{t}_{33}^{2}\geq\bar{t}_{11}^{2},\bar{t}_{22}^{2}$.
The local states are determined by the corresponding
vectors $\bar{m}_{l}$ and $\bar{n}_{l}$. Any mixed state
$\rho_{SM}$ can be uniquely converted to a state
$\bar{\rho}_{SM}$ using appropriate local unitary
operations \cite{Horodecki96b}. Thus, just two orderings of
the diagonal elements,
$\bar{t}_{11}^{2}\geq\bar{t}_{22}^{2}$ or
$\bar{t}_{11}^{2}\leq\bar{t}_{22}^{2}$, remain to be
discussed.

Let us suppose the measurements $\bar{\Pi}_{S}$ and
$\bar{\Pi}_{M}$ are constructed from projectors to the
vectors of the local bases in which $|\bar{t}_{33}|$ is
maximal. Then ${\bf \Delta
\bar{D}}=\max(0,|\bar{t}_{33}|-|\bar{n}_{3}|)$. For
$\bar{t}_{11}^{2}\geq\bar{t}_{22}^{2}$, let the
measurements $\bar{\Pi}'_{S}$ and $\bar{\Pi}'_{M}$ be
related to the bases in which $|\bar{t}_{11}|$ is maximal
so that ${\bf \Delta
\bar{D}'}=\max(0,|\bar{t}_{11}|-|\bar{n}_{1}|)$.
Simultaneously, we express the violation of any Bell's
inequalities employing the criterion from
Ref.~\cite{Horodecki95}: A state $\bar{\rho}_{SM}$ violates
Bell's inequalities if its maximal Bell factor
$B_{\mathrm{max}}=2\sqrt{\bar{t}_{11}^{2}+\bar{t}_{33}^{2}}$
lies in the interval $(2,2\sqrt{2}]$ (notice also that
$B_{\mathrm{max}}$ is invariant under local unitary
transformations). Analogical results can be derived for
$\bar{t}_{11}^{2}\leq\bar{t}_{22}^{2}$: ${\bf \Delta
D'}=\max(0,|\bar{t}_{22}|-|\bar{n}_{2}|)$. Maximal Bell
factor is then
$B_{\mathrm{max}}=2\sqrt{\bar{t}_{22}^{2}+\bar{t}_{33}^{2}}$.
Finally we obtain an inequality ${\bf \Delta
\bar{D}}^{2}+{\bf \Delta
\bar{D}'}^{2}\leq\left(B_{\mathrm{max}}/2\right)^{2}$ valid
for an arbitrary state $\bar{\rho}_{SM}$. The equality
occurs for states with zero a-priori knowledges. For such
states a non-zero knowledge can be obtained only though the
measurement on $M$.

Now we generalize these results to any state $\rho_{SM}$ as
well as for arbitrary measurements
$\Pi_{S},\Pi'_{S},\Pi_{M},\Pi'_{M}$, where
$\Pi_{S},\Pi'_{S}$ are complementary measurements. As
pointed out, any mixed two-qubit state can be uniquely
prepared from some state $\bar{\rho}_{SM}$ (of a special
form discussed above) by appropriate local unitary
transformations $U_{S},U_{M}$ acting on qubits $S$ and $M$,
respectively. Further, the transformation of the above
chosen measurements $\bar{\Pi}_{S}$ and $\bar{\Pi'}_{S}$ to
arbitrary (but still complementary) measurements $\Pi_{S}$
and $\Pi'_{S}$ corresponds effectively to the extra local
unitary transformation $U_\Pi$ acting on the qubit $S$.
Since distinguishabilities ${\bf \Delta D}(\Pi_{S})$ and
${\bf \Delta D}(\Pi'_{S})$ are invariant under any local
unitary transformation on the qubit $M$, it is sufficient
to take into account only a joint unitary transformation
$\tilde{U}_{S}=U_\Pi U_{S}$ acting on qubit $S$. For any
unitary transformations $U$ there is a unique rotation $O$
such that
$U(\vec{n}\cdot\vec{\sigma})U^{\dag}=(O\vec{n})\cdot\vec{\sigma}$.
If a state $\bar{\rho}_{SM}$ with diagonal $\bar{T}$ is
subjected to the $U_{S}\otimes U_{M}$ transformation its
correlation matrix transforms as follows
$T=O_{S}\bar{T}O_{M}^{\dag}$ \cite{Horodecki96b}. Thus a
joint unitary transformation $\tilde{U}_{S}$ can be
represented as a transformation of the correlation tensor
$T=O_{S}\bar{T}$, where $O_{S}$ is a matrix of rotation in
$R^{3}$ space.

First, we will explicitly calculate ${\bf \Delta
D}(\Pi_{S})$ and ${\bf \Delta D}(\Pi'_{S})$ for any mixed
state using the transformation $T=O_{S}\bar{T}$. Assuming
$\bar{t}_{11}^{2}\geq\bar{t}_{22}^{2}$ we obtain ${\bf
\Delta
D}(\Pi_{S})=\max(0,\sqrt{t^{2}_{33}+t_{32}^{2}+t_{31}^{2}}-|n_{3}|)$
and ${\bf \Delta
D}(\Pi'_{S})=\max(0,\sqrt{t^{2}_{11}+t_{12}^{2}+t_{13}^{2}}-|n_{1}|)$.
Then we straightforwardly get ${\bf \Delta
D}^{2}(\Pi_{S})+{\bf \Delta
D}^{2}(\Pi'_{S})\leq\left(B_{\mathrm{max}}/2\right)^{2}$.
By analogous calculations we obtain the same result for
$\bar{t}_{11}^{2}\leq\bar{t}_{22}^{2}$. Finally, since
${\bf \Delta K}(\Pi_{M}\rightarrow\Pi_{S})\leq {\bf \Delta
D}(\Pi_{S})$ and ${\bf \Delta
K}(\Pi'_{M}\rightarrow\Pi'_{S})\leq {\bf \Delta
D}(\Pi'_{S})$ we can conclude that
\begin{equation}
\label{ineq1} {\bf \Delta
K}^{2}(\Pi_{M}\rightarrow\Pi_{S})+{\bf \Delta
K}^{2}(\Pi'_{M}\rightarrow\Pi'_{S})\leq
\left(\frac{B_{\mathrm{max}}}{2}\right)^{2}\!.
\end{equation}
Thus the maximal Bell factor represents a non-trivial bound
on the sum of the squares of knowledge excesses which can
be extracted from a pair of measurements on the ``meter''
qubit. Assuming $\Pi_{M}=\Pi'_{M}$ we can also derive an
inequality analogous to that given in Ref.~\cite{Englert}:
${\bf \Delta K}^{2}(\Pi_{M}\rightarrow\Pi_{S})+{\bf \Delta
K}^{2}(\Pi_{M}\rightarrow\Pi'_{S})\leq 1$. Our analysis
shows that for $\Pi_{M} \ne \Pi'_{M}$ the unit value on the
right-hand side may be overstepped. Note also that
$(B_{\mathrm{max}}/2)^{2}>1$ only if the state violates
Bell inequalities. For details of the proofs see
Ref.~\cite{mamut}.

A natural question is how inequality (\ref{ineq1}) can be
saturated. For the class of states with vanishing a-priori
knowledges for any measurements $\Pi_{S},\Pi'_{S}$ it can
be saturated just by the appropriate choice of measurements
$\Pi_{S},\Pi'_{S},\Pi_{M},\Pi'_{M}$. In fact, it
corresponds to the transformation of the given state to the
state with diagonal correlation tensor. It was recently
shown that there are such unique local (stochastically
reversible) filtering operations $F_{S},F_{M}$ applicable
on a single copy of a qubit pair ($F_{S}^{\dag}F_{S}\leq
1_{S}$ and $F_{M}^{\dag}F_{M}\leq 1_{M}$) that transform
(with a non-zero probability) any two-qubit mixed state
into a state which is (i) diagonal in Bell basis and (ii)
has the Bell factor $B'_{\mathrm{max}}\geq
B_{\mathrm{max}}$ \cite{Verstraete02}. Since these
Bell-diagonal states have the both local states maximally
disordered the a-priori knowledges vanish. Thus -- because
the inequality (\ref{ineq1}) is satisfied also after the
filtering -- we can always saturate it with the upper bound
given by $B'_{\mathrm{max}}$ just by an appropriate choice
of the measurements $\Pi_{S},\Pi'_{S},\Pi_{M},\Pi'_{M}$
after the appropriate local filtering.



\begin{figure}
  \begin{center}
    \smallskip
  \resizebox{0.8\hsize}{!}{\includegraphics*{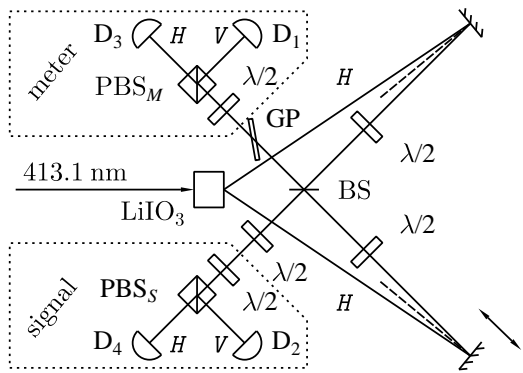}}
    \smallskip
  \end{center}
  \caption{Experimental setup.}
  \label{schema}
\end{figure}

We have verified inequality (\ref{ineq1}) experimentally
for two Werner states of qubits, $ p |\Psi^{-}
\rangle\!\langle \Psi^{-}| + \frac{1-p}{4} \openone$ (each
qubit was represented by a polarization of a photon)
\cite{Werner}. The parameter of the first Werner state
($p_1 \approx 0.82$) has been chosen so that the state was
entangled and violated Bell inequalities, the parameter of
the second one ($p_2 \approx 0.45$) so that it was
entangled but did not violate Bell inequalities.
The scheme of our experimental setup is shown in
Fig.~\ref{schema}. A krypton-ion cw laser (413.1~nm, 90 mW)
is used to pump a 10-mm-long LiIO$_3$ nonlinear crystal cut
for degenerate type-I parametric downconversion. We exploit
the fact that the pairs of photons generated by spontaneous
parametric downconversion (SPDC) manifest tight time
correlations. In our setup the photons produced by SPDC
have horizontal linear polarizations. Different
linear-polarization states are prepared by means of
half-wave plates ($\lambda/2$). The two photons impinge on
two input ports of a beamsplitter (BS) forming a
Hong-Ou-Mandel (HOM) interferometer \cite{Mandel}. A
scanning mirror is used in one interferometer arm in order
to balance the length of both arms, as indicated by an
arrow in Fig.~\ref{schema}. A glass plate (GP), that
introduces polarization dependent losses, serves to
compensate a non-ideal splitting ratio of the
beam-splitting cube (it is about 51:49 for vertical and
55:45 for horizontal polarization). HOM interferometer
enables us to prepare conditionally polarization singlet
states (i.e., $|\Psi^{-} \rangle$ Bell states).
The simplest theoretical model of the beamsplitter leads to
the conclusion that if one fetches Bell states at the input
the only one of them that results in a coincident detection
at two different outputs of the beamsplitter is the singlet
state $|\Psi^{-} \rangle$. However, in case of a ``real''
beam-splitting cube one must take into account that the two
photons strike upon a beamsplitter in \emph{opposite}
directions. So, the mutual phase (at the interface plane)
of the horizontal components of the electric-field vectors
from the two opposite inputs is shifted by $180^\circ$ just
for \emph{geometrical} reasons. Therefore it is the triplet
state $|\Psi^{+} \rangle$ that leads to a coincident
detection at different outputs. However, it is easy to
change $|\Psi^{+} \rangle$ to $|\Psi^{-} \rangle$ by means
of a half-wave plate placed in one output arm of the BS.

The mesurement block in each output arm consists of a
half-wave plate and polarizing beamsplitter (PBS). It
enables measurement in any linear-polarization basis.
Behind the PBS the beams are filtered by cut-off filters
and fed into multi-mode optical fibers leading to detectors
D$_1$,\ldots,D$_4$ (Perkin-Elmer single-photon counting
modules; quantum efficiency $\eta\approx 50\,\%$, dark
counts about 100\,s$^{-1}$).


The Werner states were prepared as a ``mixture'' of three
kinds of inputs. First we measured coincidences with
horizontal and vertical polarizations in the individual
inputs of HOM interferometer (measurement time for each
point in the following graphs was 22~s), then we added the
results of measurement with two horizontally polarized
input photons (this measurement period took 10~s), and
finally we measured with two vertically polarized input
photons (13~s). The different times of measurement
compensated the influence of a glass plate (GP) for the
vertical-vertical and horizontal-horizontal input
polarizations.
The different values of parameter $p$ were obtained
changing the position of the scanning mirror. Namely, we
have measured at $0\,\mu$m and $30\,\mu$m from the dip
center.

The measurement $\Pi_M$ on the ``meter'' qubit was
represented by a measurement in different linear
polarization bases parametrized by an angle $\vartheta$: $
\Pi_M = \{ \Pi_{M}^{+}, \Pi_{M}^{-} \} \equiv \{ |\psi
\rangle\!\langle \psi |, |\psi_{\perp} \rangle\!\langle
\psi_{\perp} | \}, $ where $| \psi \rangle = \cos \vartheta
\, | H \rangle + \sin \vartheta \, | V \rangle$ and $|
\psi_{\perp} \rangle = \sin \vartheta \, | H \rangle - \cos
\vartheta \, | V \rangle$. The angle $\vartheta$ was set by
a properly rotated half-wave plate.
Similarly, two measurements on the ``signal'' qubit,
$\Pi_S$ and $\Pi'_S$, were represented by polarization
measurements in two bases rotated by $45^{\circ}$: $ \Pi_S
= \{ \Pi_{S}^{+}, \Pi_{S}^{-} \} \equiv \{ |H
\rangle\!\langle H|, |V \rangle\!\langle V| \}, \Pi'_S = \{
{\Pi'}_{S}^{+}, {\Pi'}_{S}^{-} \} \equiv \{ |X
\rangle\!\langle X|, |Y \rangle\!\langle Y| \}, $ where $|
X \rangle = ( | H \rangle + | V \rangle ) / \sqrt{2}$ and
$| Y \rangle = ( | H \rangle - | V \rangle ) / \sqrt{2}$.
In practice we measured coincidence rates between outputs
$\Pi_{M}^{+}$ and $\Pi_{S}^{+}$ (it is denoted $C^{++}$),
between $\Pi_{M}^{+}$ and $\Pi_{S}^{-}$ (it is denoted
$C^{+-}$), etc.\ (the first sign concerns the $M$-qubit,
the second one the $S$-qubit).
Then the knowledge $\mathbf{K}(\vartheta) |\mathrm{Tr}_{MS} \Pi_{M}^{+} ( \Pi_{S}^{+} - \Pi_{S}^{-} )
\rho | + |\mathrm{Tr}_{MS} \Pi_{M}^{-} ( \Pi_{S}^{+} -
\Pi_{S}^{-} ) \rho |$ can be calculated from measured rates
as follows
\begin{equation}
  \frac{|C^{++} - C^{+-}| + |C^{-+} - C^{--}|}{C^{++} + C^{+-} + C^{-+} + C^{--}}.
 \label{exp-K}
\end{equation}
Analogously the a-priori knowledge $\mathbf{P} |\mathrm{Tr}_{MS} (\Pi_{S}^{+} - \Pi_{S}^{-}) \rho |$ can
be obtained as
\begin{equation}
 \frac{|(C^{++} + C^{-+}) - (C^{+-} + C^{--})|}{C^{++} + C^{+-} + C^{-+} + C^{--}}
 \label{exp-P}
\end{equation}
The knowledge excess is given as
$
  \mathbf{\Delta K}(\vartheta) = \mathbf{K}(\vartheta) - \mathbf{P}.
$ The quantities $\mathbf{K}'(\vartheta)$, $\mathbf{P}'$,
and $\mathbf{\Delta K'}(\vartheta)$ are obtained in the
same way just with $\Pi'_S$ instead of $\Pi_S$.
The maximal violation of Bell inequalities,
$B_{\mathrm{max}}$, can be obtained by measuring
correlation functions for two different polarization bases
on each side. Namely, for the Werner states one can choose these bases
rotated by $22.5^{\circ}$ and $67.5^{\circ}$ (with respect
to the vertical axis) on the one side and $45^{\circ}$ and
$0^{\circ}$ on the other side:
\begin{eqnarray}
 B_{\mathrm{max}} &=& |C(22.5^{\circ}, 45^{\circ})
                   + C(67.5^{\circ}, 45^{\circ})
                   \nonumber \\
                  &+& C(22.5^{\circ}, 0^{\circ})
                   - C(67.5^{\circ}, 0^{\circ})|,
 \label{exp-B}
\end{eqnarray}
where the correlation function $C(\vartheta_{1},
\vartheta_{2})$ is estimated from the measured data as
\begin{equation}
  \frac{C^{++} + C^{--} - C^{+-} - C^{-+}}{C^{++} + C^{+-} + C^{-+} +
  C^{--}}.
 \label{exp-corr}
\end{equation}
Let us note that for Werner states the theoretical
predictions of regarded quantities read:
$
 \mathbf{K} =  p \, |\cos (2 \vartheta)|,\;
 \mathbf{K}' =  p \, |\sin (2 \vartheta)|,\;
 \mathbf{P} = \mathbf{P}' =  0,\;
 B_{\mathrm{max}} =  p \, 2 \sqrt{2}.
$
Clearly, maximal value of $\mathbf{\Delta K}^2(\vartheta)
+ \mathbf{\Delta K'}^2(\vartheta')$ should appear for
$\vartheta=0^\circ$ (and $90^\circ$),
$\vartheta'=45^\circ$.

\begin{figure}
  \resizebox{0.8\hsize}{!}{\rotatebox{-90}{\includegraphics*{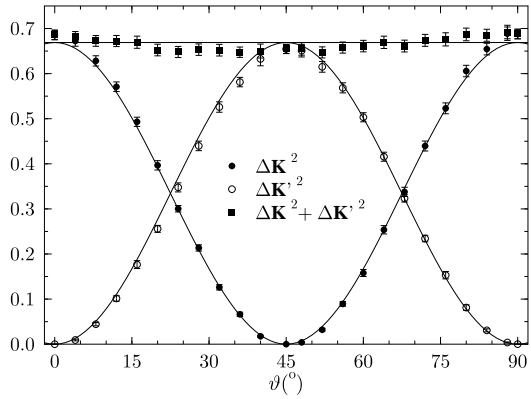}}}
  \caption{The squares of the
knowledge excesses and their sum measured
for the Werner state with $p \approx 0.82$. Symbols show
experimental values, full lines theoretical predictions
(for $p = 0.82$).}
  \label{0.8-1}
\end{figure}
\begin{figure}
  \resizebox{0.8\hsize}{!}{\rotatebox{-90}{\includegraphics*{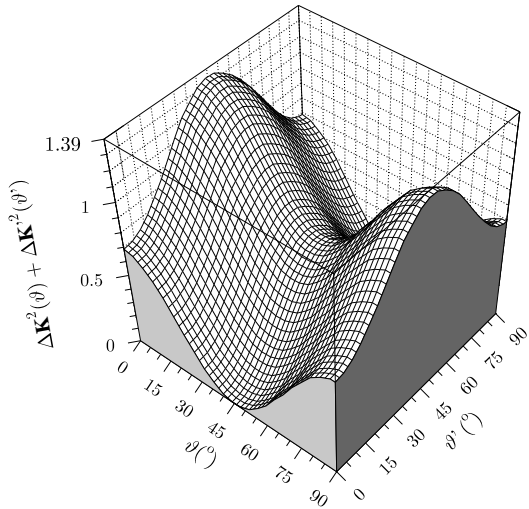}}}
  \caption{The measured values of the sum $\mathbf{\Delta K}^2(\vartheta) +
\mathbf{\Delta K'}^2(\vartheta')$ as a function of two
angle variables for the Werner state with $p
\approx 0.82$. The maximal displayed value of the vertical
axis shows the measured value of $(B_{\mathrm{max}}/2)^2$.}
  \label{0.8-B}
\end{figure}
\begin{figure}
  \resizebox{0.8\hsize}{!}{\rotatebox{-90}{\includegraphics*{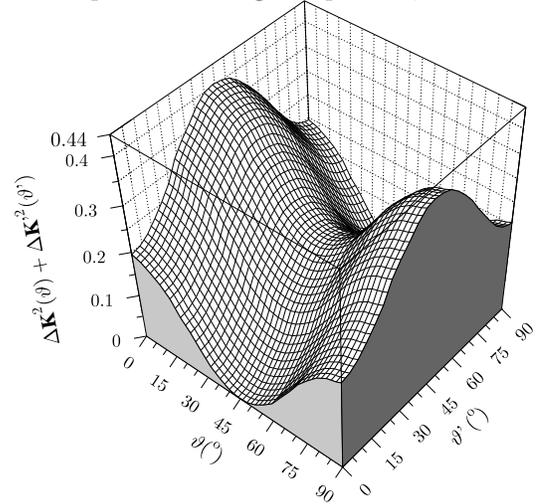}}}
  \caption{The measured values of the sum $\mathbf{\Delta K}^2(\vartheta) +
\mathbf{\Delta K'}^2(\vartheta')$ as a function of two
angle variables for the Werner state with $p
\approx 0.45$. The maximal displayed value of the vertical
axis shows the measured value of $(B_{\mathrm{max}}/2)^2$.}
  \label{0.5-B}
\end{figure}

The following graphs display our experimental results. In
Fig.~\ref{0.8-1} there are plotted the squares of the
knowledge excesses $\mathbf{\Delta K}^2(\vartheta),
\mathbf{\Delta K'}^2(\vartheta)$ and their sum measured for
the Werner state with parameter $p \approx 0.82$ (this
parameter was estimated from the best fit accordingly to
the theoretical predictions for Werner states). The error
bars show statistical errors. The accuracy of
polarization-angle settings was about $\pm 1^\circ$.
Fig.~\ref{0.8-B} shows the sum $\mathbf{\Delta
K}^2(\vartheta) + \mathbf{\Delta K'}^2(\vartheta')$ as a
function of two angle variables for the same Werner state.
The maximal displayed value of the vertical axis determines
the measured value of $(B_{\mathrm{max}}/2)^2$. The maximal
measured Bell factor is $B_{\mathrm{max}} = 2.36 \pm 0.02$
what is in a good agreement with the theoretical value for
$p=0.82$ that equals 2.319.
%
%
%
%
The same kind of measurement is presented in
Fig.~\ref{0.5-B} but now for the Werner state with $p
\approx 0.45$. The corresponding measured maximal Bell
factor is $B_{\mathrm{max}} = 1.32 \pm 0.02$ (theoretical
value for $p=0.45$ is 1.273). As can be seen, for the both
measured states the experiment has verified inequality
(\ref{ineq1}).
%
%
%
%


The measurement on the one of two correlated particles give
us a power of prediction of the measurement results on the
other one. Of course, one can never predict exactly the
results of two complementary measurements at once. However,
knowing what kind of measurement we want to predict on
``signal'' particle, we can choose the optimal measurement
on the ``meter'' particle. But there is still a fundamental
limitation given by the sort and amount of correlations
between the particles. Both these kinds of constraints are
quantitatively expressed by our inequality.


\medskip
\noindent {\bf Acknowledgments.} We would like to thank
J.~Fiur\'{a}\v{s}ek and L.~Mi\v{s}ta,~Jr. for fruitful
discussions. This research was supported by the projects
202/03/D239 of GACR and LN00A015 and CEZ: J14/98 of the
Ministry of Education of the Czech Republic.


\end{document}